# Disks, Partitions, Volumes and RAID Performance with the Linux Operating System

**Michel R. Dagenais (michel.dagenais@polymtl.ca)**
Dept. of Computer Engineering, Ecole Polytechnique de Montréal, P.O. Box 6079, Station Centre-Ville, Montreal, (Quebec), Canada, H3C 3A7

*Abstract*

Block devices in computer operating systems typically correspond to disks or disk partitions, and are used to store files in a filesystem. Disks are not the only real or virtual device which adhere to the *block accessible stream of bytes* block device model. Files, remote devices, or even RAM may be used as a virtual disks. This article examines several common combinations of block device layers used as virtual disks in the Linux operating system: disk partitions, loopback files, software RAID, Logical Volume Manager, and Network Block Devices. It measures their relative performance using different filesystems: Ext2, Ext3, ReiserFS, JFS, XFS,NFS.

## Introduction

As Central Processing Units (CPU) rush into the multi Gigahertz frequency range, with several instructions issued per clock cycle, disks have long been predicted to be the weak link of the performance chain [1]. Fortunately, computer architects have concentrated on this weak link and file access performance has kept up with the CPU performance better than suggested by the disk seek time alone.

Not only are disks a crucial aspect of performance, this is where all the persistent information is stored. Their organisation strongly impacts the maintenance required (e.g. moving files when a disk is full, recovering data upon disk failure) and the survivability to crashes (e.g. accidental reboot). A number of block device layers may be used to obtain a more flexible organisation than one device per physical disk [2] [3] (e.g. /dev/hda and /dev/hdc for the first disk of each of the two IDE chains typically available on a personnal computer...).

- Disk partitions may be used to access a single large disk as several separate smaller disks (e.g. partitions within /dev/hda are /dev/hda1, /dev/hda2...). These smaller logical disks (partitions) may isolate different sections of the file hierarchy (system files versus user files), or different operating systems and versions.

- A virtual disk in a file, (loop device), provides the same level of isolation. These are much easier to create than disk partitions but less efficient and are thus commonly used for simple experimentations. Unlike real disks or partitions, a file may happen to be stored in non contiguous blocks on the physical disk.

- Software RAID block devices is a layer above regular block devices to redirect block accesses [4]. In RAID 1 mode, the RAID block device (e.g. /dev/md0) is layered over two block devices of the same size (e.g. /dev/hda1 and /dev/hdc1) and keeps the two mirrored. Each block written to the virtual device /dev/md0 is thus converted to writes to both /dev/hda1 and /dev/hdc1. A read from the same /dev/md0 is converted to a read from the least busy of /dev/hda1 and /dev/hdc1, since they have the same content. RAID 1 is typically used to mask and tolerate disk failures.

- Logical Volume Management is a layer above regular block devices to separate logical partitions from physical partitions [5]. This can be used to move a logical partition from one disk to another (disk replacement) without bringing it offline. Logical partitions may also span several disks, allowing to increase a partition size by simply adding a new disk, adding it to a logical volume and enlarging the virtual partition.

- Network Block Devices are proxys for devices on remote computers [6]. They may be used for diskless computers or for RAID mirroring across the network.

These block device layers may be combined in different useful ways. Logical Volumes are often built on top of Software RAID devices. Network Block Devices, when used by numerous diskless clients, frequently access loop devices on the server instead of requiring one partition per client.

Network Block Devices may also be used as a remote disk in a RAID 1 mirror. For instance, an important computer may use a RAID 1 mirror between its local disk and a remote backup file (Network Block Device to a Loop Device to a file). When this computer fails (e.g. gets stolen or hit by a bus), a replacement computer may be brought online very quickly using the remote backup copy as the correct device in a degraded RAID 1 mirror. The empty local disk of the replacement computer is added to the degraded RAID 1 mirror as a replacement disk to gradually resynchronize. The replacement computer is thus immediately operational in degraded mode (remote disk and no redundancy) and fully operational after an hour or so, once the mirror resynchronisation is completed.

The layer between block devices and a file hierarchy is the filesystem. Filesystems determine the layout of files, directories, and related information on a block device. The development of an efficient filesystem presents challenges similar to dynamic memory allocation and searching structures, with the added complexity of maintaining database like atomicity in case of system crashes. Furthermore, the access time of blocks, and thus the performance, is strongly dependent on the associated disk rotation and head movement. This complex problem with a vast solution space explains why several high quality filesystems are currently competing for a place on your disk:

- Ext2 won the filesystem competition in the early 0.x to 2.2 Linux era. It is robust and efficient. If the system crashes in the middle of a filesystem modification, corruption may occur (e.g. blocks from a removed file have been removed from the file inode but not yet added to the free blocks list). When rebooting from an unplanned shutdown, a full filesystem check is therefore always performed. Unless explicitly mentioned otherwise, this is the default filesystem used in the tests presented in later sections.

- Ext3 is a compatible extension to Ext2 which adds journaling. In case of a system crash, the journal describes the planned operations. Upon reboot, the filesystem checks the journal to determine which planned operations were completed and which need to be completed. Lengthy filesystem checks are thus avoided.

- ReiserFS is a newer journaled filesystem with optimisations for numerous small files and large directories. It was the first journaled filesystem incorporated into the Linux kernel.

- JFS is a journaled filesystem initially developed at IBM and suitable for large servers (numerous files, very large files...).

- XFS is a journaled filesystem initially developed at SGI, also targeted at large servers.

- NFS is not a native filesystem but rather a Network File System layered over a regular filesystem. The remote operations are performed at the files and directories level, instead of at the block level in the Network Block Device. Unlike Network Block Devices, NFS files may be accessed from several hosts simultaneously.

In the context of a low maintenance, thin client fat server, architecture targeted at elementary schools, several different server configurations were being considered. Thin diskless clients could use NFS or Network Block Devices (on a loopback device) to access their root filesystem. Servers had a choice of several filesystems and could use logical devices, software RAID, and RAID over Network Block Devices. This motivated the current evaluation of the performance impact of each of these alternatives.

The next section details the performance test setup. Following sections each study a specific parameter which may impact performance: disk size, partition size and location, filesystem type, block device in a file loopback device, software RAID, logical volumes, and Network Block Devices. The article ends with a discussion of the main results.

## Experimental Setup

The tests were performed on a 900MHz Pentium 3 computer with 512MB RAM running Linux version 2.4.8-26 as distributed by Mandrake. The root filesystem was read from a floppy disk upon booting and kept in a small RAM disk. Therefore, the IDE disks were strictly used for running the tests. Four different input-ouput test programs were used:

- Bonnie++, a synthetic benchmark which measures specific parameters such as file creation, write, read and removal times [7],
- Dbench, a realistic benchmark which recreates the trace of file operations seen by a Samba file server [8],
- Linux kernel compilation, a recompilation of the Linux kernel source code,
- Tar, creation of an archive file for the kernel source tree.

The Dbench benchmark is probably the best indication of the overall performance to be expected from a server disk. Interestingly, disk writes dominate in this benchmark, which may be surprising since the operating system normally receives about ten times more read than write through system calls issued by processes. However, read operations are very often served directly from buffer caches in RAM, both in the client and in the server. Thus, especially with the amount of RAM now routinely found in computers, relatively few read operations are actually performed by the disks (typically the first access since boot time to a file).

The results in the various tests have been converted to units directly proportional to bytes per second, for instance the total bytes read and written divided by the elapsed time for the Tar test. Thus, in all the results presented, more is better. For Dbench results, the output is already in million bytes per second and has been used directly. For Bonnie++, the sequential block reads value in million bytes per second is presented.

## Disk capacity

The total capacity is rarely the only difference between two disk drives. Larger capacity disks are typically more recent, while retaining the same form factor and number of platters. The increase in tracks and bits density explains the increased capacity. Even if the head movement and disk rotation speeds are unchanged, the disk head transfer rate and track to track speed may increase because there are more bits on a given track and tracks are more closely spaced.

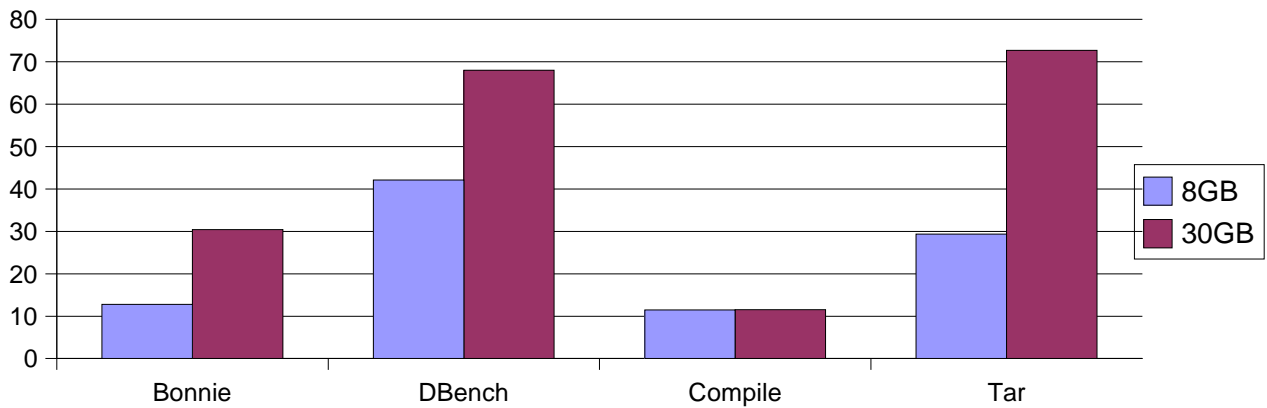
*Figure 1: Relative performance of a small and a larger disk.*

The speed difference between the older smaller disk and the newer larger disk is very significant, much larger than the almost negligible difference in average seek time. Similar results were obtained with 2GB and 100GB disks. The Linux kernel compilation time is largely unaffected by disk performance being CPU bound.

With the amount of RAM now available for the buffer cache, operating systems have been optimized to perform much fewer but considerably larger disk accesses through aggressive disk read ahead (reading a full track when a sector is needed) and write combining (waiting for several contiguous blocks to write within a few seconds before actually writing). Therefore, the sequential read and write speed is becoming the dominating factor of disk performance. The increased linear bit density along each track thus explains the much better performance of larger capacity disks. At some point, the interface between the disk and the computer may become a bottleneck. This is being actively worked on by system architects with faster versions of IDE, including the more recent SATA.

## Partition Size and Location

Few system administration books, if any, discuss the size and placement of disk partitions in terms of performance. Other considerations such as isolation and flexibility for reinstallation are often mentioned. A disk was partitioned in several ways to measure the effect of partition size (at the same location) and partition location (with the same size at different locations).

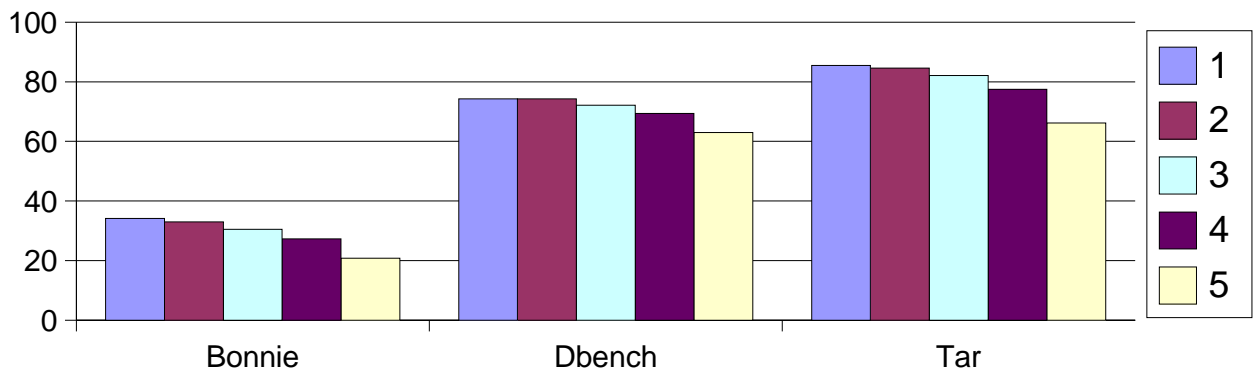
*Figure 2: Disk performance for several 2GB partitions with 7GB gaps between them. Partition 1 is /dev/hda1, partition 2 is /dev/hda3...*

Interestingly, there is a measurable performance difference between the different partition locations. While disk manifacturers do not specify how blocks are positioned on the disks, some SCSI disks allow querying the number of zones and the number of sectors per track in each zone. The data obtained in this way suggests that magnetic disks, unlike CDROMs, usually have their tracks numbered from the outside towards the inside. Thus, the 2GB /dev/hda1 partition is on the outer edge, and contains fewer tracks than /dev/hda3, each track containing more bits, for a constant linear density but a faster head transfer rate, since the head to disk speed increases with the radius. The results obtained here demonstrate that this is clearly the case.

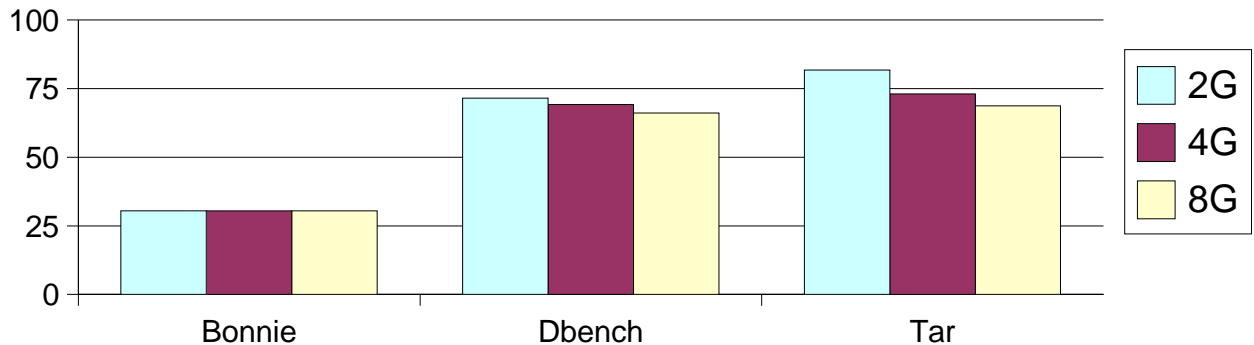

*Figure 3: Disk performance for partitions of different sizes.*

The effect of partition size is less obvious and is probably related to location. The 2, 4 and 8GB partitions tested all have the same starting location but extend farther for the 4 and 8 GB partitions. Reading or writing blocks at the beginning of any of these partitions should take the same time. However, if files are located by default close to the middle of the partition, the middle of the 2GB partition is closer to the disk edge than the middle of the 8GB partition and performs slightly better. However, the difference in that case should be smaller than between partition 1 and partition 2 in Figure 2 which are 7GB apart, which is not the case. There must be more seeking involved in larger partitions, explaining the lower performance, either because the files are spread over the larger partition, or because other elements such as inodes or replacements for defective blocks are farther apart from the rest of the blocks allocated to the test files.

## Filesystems

Filesystems must support a wide range of file operations (creation, deletion, read, write, directory access...), and widely different data structures on disk may optimize some operations at the expense of others. The sequential read operation in Bonnie++ performs equally well in all filesystems. Dbench performance varies significantly from one filesystem to the other with its mix of large and small read and write issued from several clients. Ext2, not carrying the burden of maintaining a journal of modifications, shines given the large number of writes. ReiserFS also performs well, probably due to its optimized directory structure.

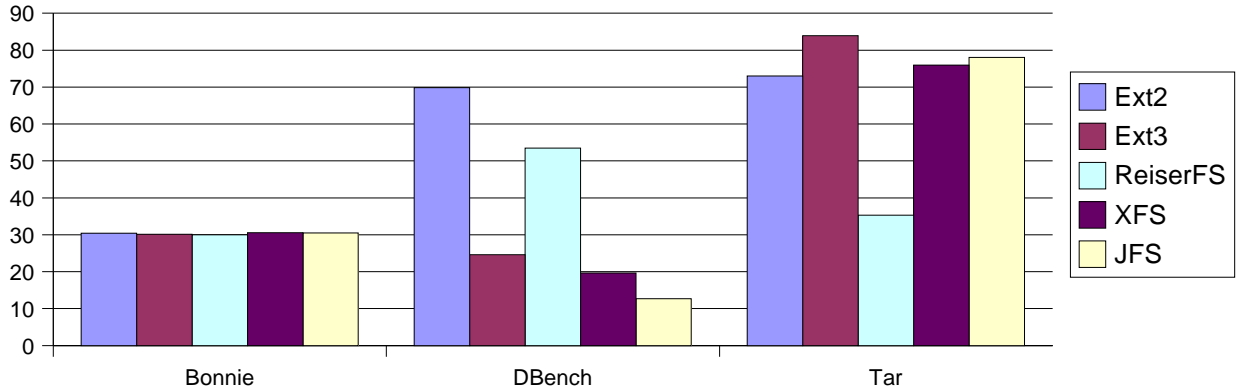
*Figure 4: Disk performance for several different filesystems.*

In the Tar benchmark, numerous small files are read and a single tar archive file is written. All filesystems perform equally well in that case, except for ReiserFS. It is important to note, however, that all these filesystems are constantly being improved upon, with the possible exception of Ext2, and these figures are changing from one kernel version to the other.

## Virtual Disk in a File

For isolation or experimentation purposes, it is often useful to use a virtual disk in a file. For example, each virtual disk may be a root device for a thin client, or a system disk for a legacy emulated operating system. Filesystems and buffer cache read ahead policies are normally optimized to insure that sequential blocks in a file are mostly contiguous on disk and can be accessed quickly. With the indirection layer of a virtual disk in a file, the operating system cannot as easily perform these optimisations.

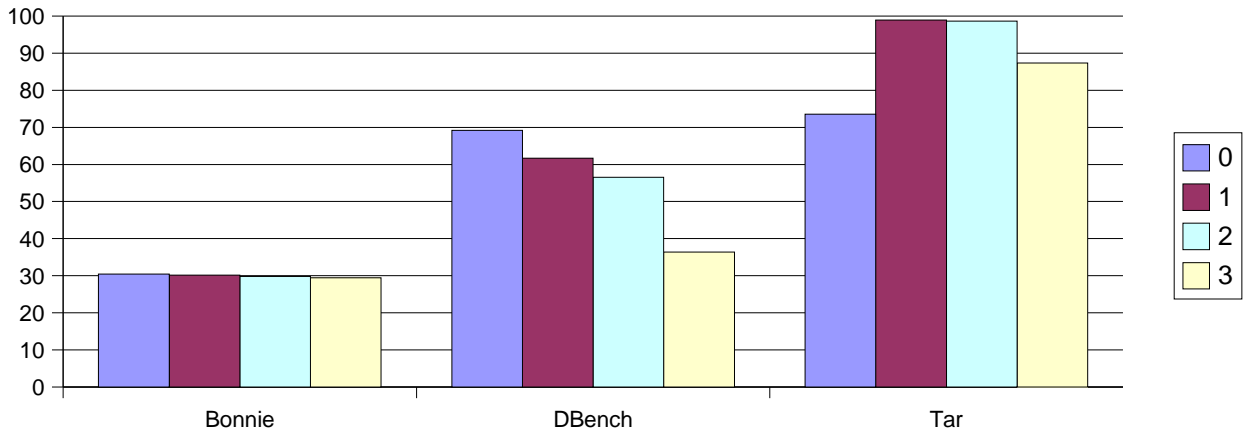
*Figure 5: Performance using none, one, two or three virtual disk indirection layers.*

The performance tests were run:

0) on the normal physical disk, /dev/hda2),

1) on a virtual disk (/dev/loop1) in file /mnt/hda2/loopfile on disk /dev/hda2,

2) on a virtual disk (/dev/loop2) in file /mnt/loop1/loopfile on virtual disk /dev/loop1,

3) on a virtual disk (/dev/loop3) in file /mnt/loop2/loopfile on virtual disk /dev/loop2.

The Bonnie and Dbench tests, as expected, show a steady decrease in performance as the number of virtual disk layers increases. The Tar test is more erratic with the physical disk being surprisingly less performing.

## Software RAID Performance

For this test, two different disks were used, Ext2-1 a more recent 30GB drive, and Ext2-2 an older 8GB disk drive. The tests were then performed with the two drives mirrored using software RAID (i.e. RAID level 1), with the Ext2 filesystem and with the Ext3 filesystem.

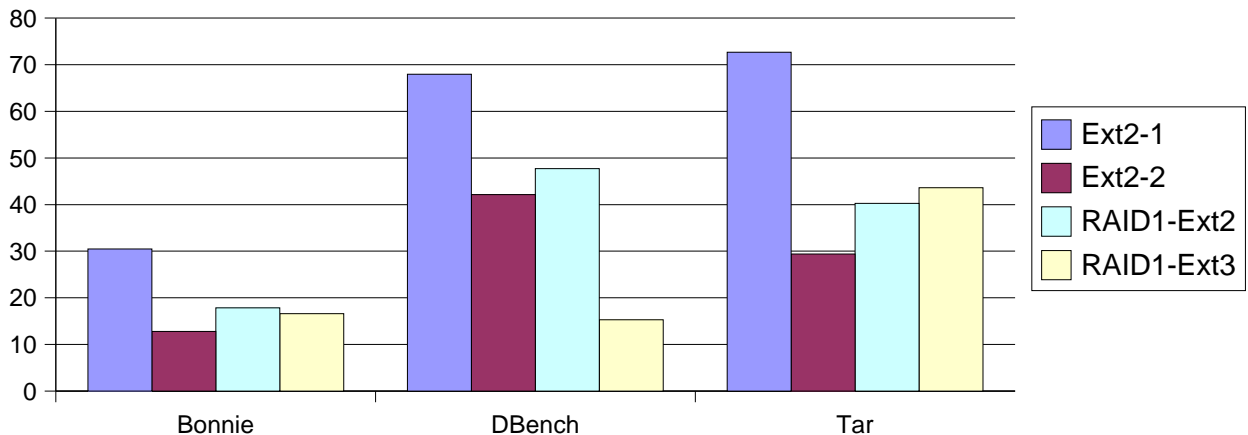

*Figure 6: Software RAID performance for a mirror built from a faster (Ext2-1) and a slower (Ext2-2) disk.*

The newer drive performs much better than the older drive. When the disks are mirrored, read operations can be performed on either one, yielding the performance of the better drive, but write operations must be performed on both, yielding the performance of the slower drive. As can be expected, the RAID 1 performance is therefore somewhat in between the performance of the two drives. As seen in a previous test, the journaling of Ext3 comes on average at some cost in performance.

## Logical Volume Management

Logical volumes add a mapping layer between the physical disks and the logical partitions. It is then possible to group several disks into a large logical volume, or even to add a disk to a logical volume and subsequently to enlarge an existing partition in this logical volume. More advanced features are also offered such as migrating space in a volume from one disk to another (to replace a disk), and instant snapshots for live backups (modifications are written to new blocks and the old content is preserved such that a frozen snapshot is retained).

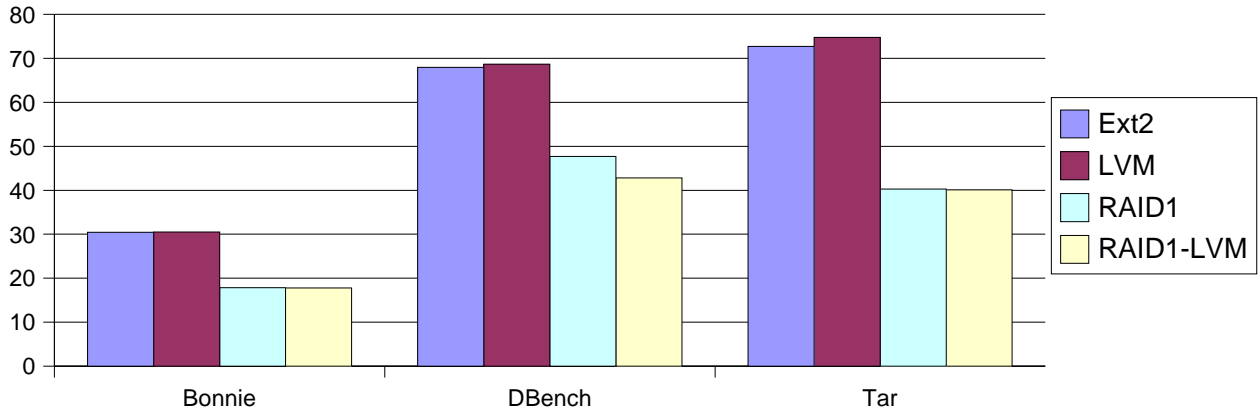

*Figure 7: Performance of a disk partition and a software RAID mirror with and without a Logical Volume Management Layer (LVM).*

The performance impact of adding a logical volume management layer is almost negligible either on top of a physical disk (Ext2 versus LVM) or on top of a software RAID 1 disk mirror (RAID 1 versus RAID 1-LVM).

## Network File Systems

The tests were run on the local disk, through the Network Block Device daemon from a local disk (only worked for the Tar test), through the Network Block Device from a remote disk, on a software RAID 1 mirror between two local disks, on a RAID 1 mirror between a local disk and a Network Bock Device remote disk, and on a NFS mounted filesystem from a remote disk.

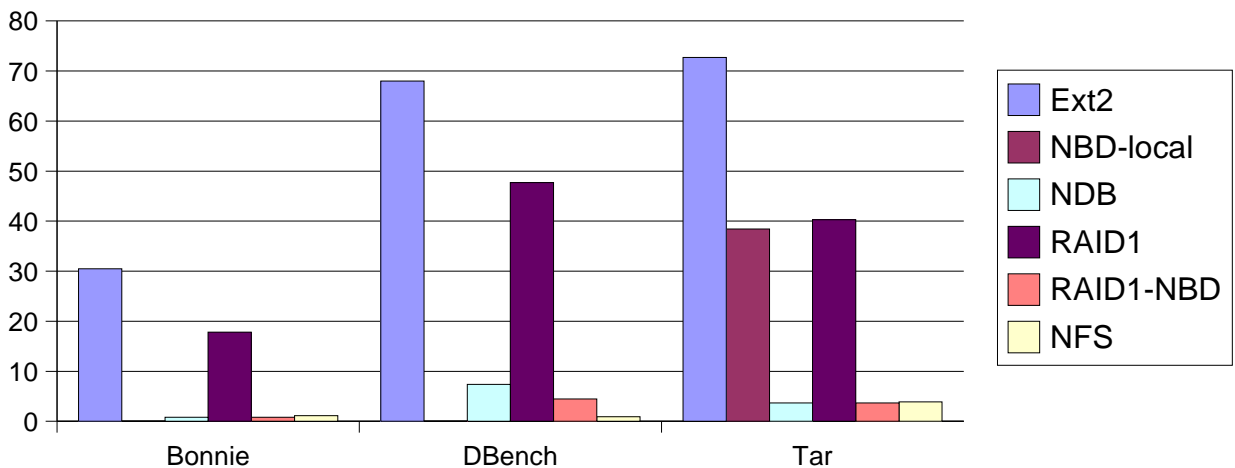

*Figure 8: Performance of a disk partition versus the same partition accessed through the Network Block Device daemon (NBD-local), a remote partition accessed by NBD, a local software RAID mirror (RAID1), a software RAID mirror using one local and one remote partition (RAID1-NBD), and a remote partition accessed through the Network File System (NFS).*

The overhead of using a network protocol and copying the blocks is significant, as can be seen in the NBD over local disk test. Moreover, the Network Block Device system has not seen widespread heavy production use, and optimisation. In this test, the network block devices and filesystems were much slower than the local disks. A portion of this decrease may be attributed to the overhead of the network protocol but the major culprit is the network itself, at 10 megabits/second (approximately 1 megabyte/second). A 100 megabits/second network (approximately 10 megabytes/second) would still be a limiting factor, while a gigabit network would have a performance comparable to that of newer IDE drives and not be as much of a bottleneck.

Remote file servers are extremely useful to share files, to provide centralized maintenance and backups, and to offer fault-tolerant distributed replication. There is, however, some overhead involved in communicating through the network. Furthermore, disk transfer rate improvements happen on a yearly basis while networking speed improvements happen every 5 to 10 years and often lag behind. The result can be a significant reduction in performance, as illustrated by the factor of 70 between the local disk and the NFS mounted filesystem performance for the DBench test. It is, therefore, important for performance sensitive applications to carefully select which files are local and which files are remote, and to enable client caching whenever possible.

## Discussion

Most of the advanced features discussed here, such as journaled filesystems, RAID and Logical Volume Management, have appeared in dedicated, expensive, factory configured servers in the 1980s. With the advent of versatile commodity servers and free software operating systems, system administrators now have to decide among several possible organizations: a choice of no less than four modern journaled high performance filesystems (Ext3, JFS, ReiserFS, XFS), software RAID, Logical Volume Management, and remote devices and filesystems.

A typical large file server will use a number of physical disks organized as software or hardware RAID 1 or RAID 5 (for redundancy and speed), used under Logical Volume Management (for live resizing and migration of logical partitions, and snapshots for backups), formatted with any of the four high performance journaled filesystems (for quick and automatic reboot after soft crashes). These filesystems will then be offered to clients through the network, insuring a very high speed connection between the server and its network switch, connecting it to the clients. While NFS version 3 is the mature file serving network protocol, NFS version 4 [9] and newer versions of SMB (implemented in Samba [10]) perform more aggressive caching and often perform much better.